# Working with Affective Computing: Exploring UK Public Perceptions of AI enabled Workplace Surveillance.

Lachlan Urquhart [0000-0001-5144-5024] Alex Laffer 2[0000-0003-2463-9135] Diana Miranda 3[0000-0002-8605-5031]

¹ University of Edinburgh, Edinburgh, Scotland
² Bangor University, Bangor, Wales
³ University of Stirling, Stirling, Scotland

lachlan.urquhart@ed.ac.uk

**Abstract.** This paper explores public perceptions around the role of affective computing in the workplace. It uses a series of design fictions with 46 UK based participants, unpacking their perspectives on the advantages and disadvantages of tracking the emotional state of workers. The scenario focuses on mundane uses of biometric sensing in a sales environment, and how this could shape management approaches with workers. The paper structure is as follows: section 1 provides a brief introduction; section 2 provides an overview of the innovative design fiction methodology; section 3 explores wider shifts around IT in the workplace; section 4 provides some legal analysis exploring emergence of AI in the workplace; and section 5 presents themes from the study data. The latter section includes discussion on concerns around functionality and accuracy of affective computing systems, and their impacts on surveillance, human agency, and worker/management interactions.

**Funding:** We would like to thank our funders for supporting this research under Economic and Social Research Council grant ES/T00696X/1 [All] and UKRI Engineering and Physical Science Research Council grants EP/T022493/1 and EP/V026607/1 [Urquhart]

# 1        Introduction: Affective Computing in the Workplace

Affective computing (AC) systems are emerging in the workplace changing the nature of workplace cultures and enabling technologically mediated professional relationships. Better understanding of employees' emotions may have benefits like protecting worker wellbeing, but significant risks are likely to emerge such as forms of tracking that benefit employers at the expense of employee interests. In this paper, we explore public perceptions of mundane uses of AC and biometric surveillance systems at work. Through a series of ten online workshops with 46 members of the UK public, we used a novel design fiction led methodology to elicit insights. We observed concerns around AC functionality and accuracy, negative impacts on human agency and interactions, and general anxieties around expanding surveillance infrastructures. To contextualise these concerns beyond our empirical findings, we present analysis of the legal dimensions of AI use at work and the wider histories of information technologies being integrated into the workplace.

# 2        Methodology: Design Fictions

To explore a range of Affective Computing (AC) use-cases in a situated manner, we developed an innovative narrative approach (Laffer 2022). This draws on **Design Fiction** (Bleecker 2009; Jensen & Vistisen 2017; Coulton et al, 2017) – notably the use of diegetic prototypes; technology that exists within a fictional world –  and aspects of **Contravision** (Mancini et al. 2010), where positive and negative outcomes for each use-case are created. A storyline narrative, created using Twine (an interactive fiction writing tool), was presented in online workshops to engage our participants (n=46) with mundane examples of AC. This supported a narrower focus on people, social practices, and technological impacts. This process of 'domestication' (Auger 2013) helped to quickly familiarise participants with socio-technical aspects of emergent systems and go beyond just the technological aspects. The full Twine can be viewed here: https://eaitwine.neocities.org/ and the opening image of the AffecTECH narrative is provided below (Figure 1).

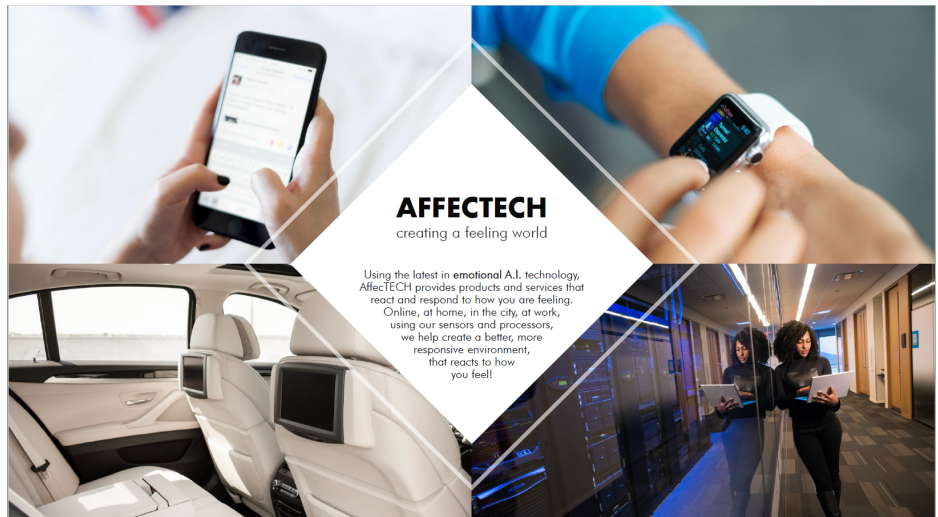

*Figure 1: Advert for AffecTech, a fictional technology company, presented to participants during online workshops as part of a Twine narrative.*

Participants were separated into three older groups (65+) (n=13); three younger groups (18-34) (n=12), two groups of disabled participants (n=10) and two groups of people belonging to UK ethnic minorities (n=11). Workshop discussions were recorded, transcribed, and uploaded to qualitative analysis software Nvivo for thematic analysis.

In our workplace scenario, a new emotion sensing system (AffecTECH) has been introduced into a call center so employees making sales calls are monitored and evaluated through voice and other biometric data. The system provides immediate performance feedback and maintains these records for use in staff evaluation by managers. The technology is introduced via a diegetic email (figure 2) from the protagonists' manager. The contravision element is composed of positive and negative feedback about the system by two of the protagonists' colleagues who had piloted the technology.

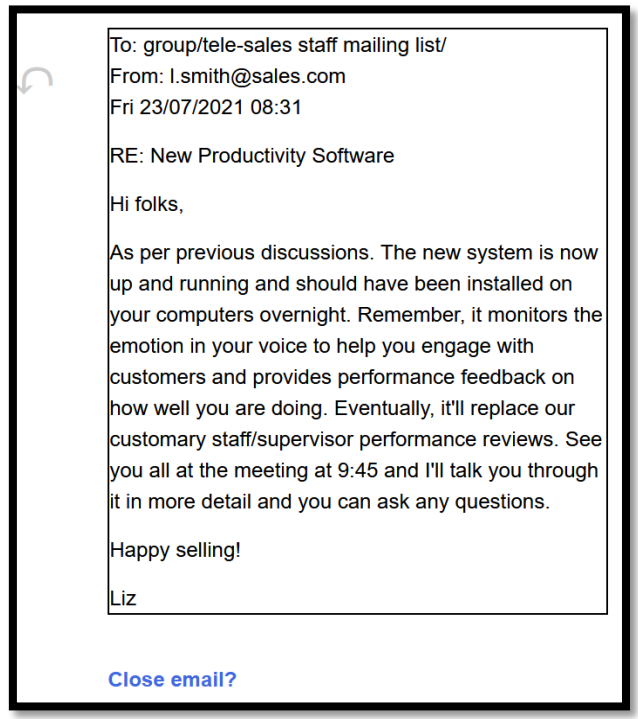

To: group/tele-sales staff mailing list/
From: l.smith@sales.com
Fri 23/07/2021 08:31

RE: New Productivity Software

Hi folks,

As per previous discussions. The new system is now up and running and should have been installed on your computers overnight. Remember, it monitors the emotion in your voice to help you engage with customers and provides performance feedback on how well you are doing. Eventually, it'll replace our customary staff/supervisor performance reviews. See you all at the meeting at 9:45 and I'll talk you through it in more detail and you can ask any questions.

Happy selling!

Liz

**Close email?**

*Figure 2: Email introducing the workplace AC use-case presented to workshop participants as part of a Twine narrative.*

## 3 Technologically Mediated Workplace Surveillance

Before presenting results, it is useful to reflect on wider literature and concerns around incorporation of new IT systems into the workplace, as this has long raised legal and design issues. For example, in Scandinavia, researchers and trade unions in the 1970's and 80's tried to shape the early emergence of IT at work to address risks like deskilling of workers tasks, and IT enabling poor management practices (Bjerknes and Bratteig, 1995). Their goal was to instill a form of **workplace democracy** where workers participated in the design of IT to ensure they would continue to flourish at work and new IT systems would not negatively impact existing working practices (Floyd, 1989). Moving forward to present day, current workplace IT incorporates data driven machine learning tools for overt or covert profiling and management of the workforce (Edwards, Martin and Henderson, 2019). As Jarrahi et al (2021) highlight, we are moving to an age of 'algorithmic management' in workplaces. Yet lessons from 40 years ago have not been learned, and many ethical and legal concerns remain. Four brief examples show this. **Gig economy platforms** such as Deliveroo or Uber track and monitor movements and performance of workers remotely, yet workers have limited agency to change the platform (Prassl, 2019). **Automated hiring systems** utilise machine learning to determine if the candidates match arbitrary psychometrics, which can in turn lead to discrimination (Monedero, Dencik, and Edwards, 2020). Some systems even claim to detect emotional states, reiterating the turn to metricising emotions (HireVue – Chen and Hao, 2020). Employer **wellness programmes** utilise employee wearables to track vital statistics (e.g., heart rate, activity) but can leave privacy concerns around data handling unaddressed (Iliadis and Pederson, 2018). Exacerbated by the shift to remote working prompted by the global Covid-19 pandemic, employers have been increasingly monitoring employees through **covert surveillance using webcams, facial recognition, and bespoke software**. This challenges trust relationships between employer and employee, raises privacy concerns, and surfaces differentiated impacts of surveillance, particularly across genders (Stark, Stanhaus and Anthony 2020). Yet, we see commercial vendors such as Uniphore offer to monitor customer sentiment to provide opportunities for sellers to add value to their calls, and build a trusting relationship with a customer[1]. Similarly, video conferencing suite Zoom, is looking to deploy emotional AI in video analytics.[2] Given the concerns around the ability of emotional AI to even detect emotions in the first place, and the resulting criticisms of it being a form of 'pseudoscience' (Crawford, 2021), the commercial plans to roll out such systems at scale in workplaces are concerning.

These shifts appear to be here to stay. Ball (2021) has highlighted that "thoughts, feelings, and physiology" are a key target in **workplace surveillance** through use of biometrics, wearables, and emotion tracking e.g., in call centres (Bromuri, Henkel, Iren and Irovi, 2020). Ball's earlier work on call centres explored negative impacts of extensive workplace surveillance (Ball and Margulis, 2011), and as monitoring of voice tone is delegated to machines in the future (McStay, 2018), our scenario draws together these trends.

## 4     Surely the Law can help?

To tackle ethical and legal challenges posed by AC, the **proposed EU AI Act**[3] may benefit workers by preventing certain AI applications from ever becoming available on the EU market in the first place. The legislation seeks to **regulate AI** in a product safety driven manner, utilising different standards depending on the levels of risk a system poses, from prohibited, to high to low risk. Hiring systems, for example, are deemed to be **high-risk AI systems** (HRAIS) in the AI Act. These are systems where AI is used for 'recruitment or selection of natural persons,  including 'advertising vacancies, screening or filtering applications, evaluating candidates in the course of interviews or tests' (Annex III, AIA). Other workplace HRAIS include AI used for making 'decisions on promotion or termination of work-related contractual relationships, for task allocation, and for monitoring performance and behaviour of persons in such relationships' (Annex III, AIA). A system being deemed a HRAIS means it faces a wide range of design and development requirements for the providers and users/deployers (McStay and Urquhart, 2022). For example, there are requirements to comply with strict data governance requirements around training datasets; providing human oversight for operators to intervene when systems go wrong; and use of conformity assessment processes to ensure systems adhere to legal requirements before being available for sale (Urquhart, Crabtree and McGarry, 2022). Yet, not all emotion sensing technologies are treated strictly. In other contexts, they are only set to be regulated as **low risk systems**, requiring operators merely to be transparent with persons that they are interacting with an AI in the first place (if it is not already obvious from the context of use) (Art 52, AIA).

---

[1] https://www.uniphore.com/

[2] https://www.protocol.com/enterprise/emotion-ai-sales-virtual-zoom

[3] https://eur-lex.europa.eu/legal-content/EN/TXT/?uri=CELEX%3A52021PC0206

EU data protection authorities are concerned that the provisions in the AIA are not strict enough. They are calling for all emotional AI systems to be treated as **prohibited** forms of AI, unless for research or health purposes (EDPS/EDPR, 2021). This would see them treated in the same way as other prohibited systems such as social crediting or automated, live, public space facial recognition by the police (see Urquhart and Miranda, 2021). Further, unlike with data protection laws such as the EU General Data Protection Regulation 2016, which provides a range of data subject rights over personal data, the proposed AI Act does not provide direct rights for individuals who may subject to AI systems (Edwards, 2022). This is concerning, given the gaps within the existing data protection and privacy law landscape workers data interests, and the AI Act will not provide direct recourse either.

A recent report from Allen and Masters (2021) examines the challenges of AI in workplace and the issues workers face in asserting control over how their data is used by employers. For example, they highlight that under European Convention on Human Rights case law there can still be a reasonable expectation to private and family life during employment (as per Barbulescu v Romania).[4] The state needs to provide appropriate legal frameworks to guard against unnecessary, disproportionate intrusions by employers, and the facts of any specific case will be key to determining if an intrusion is justifiable or not. Under data protection law even in employment relationships, a **lawful basis** is needed for processing data (Art 6, GDPR), which guards against employers acting in uninhibited ways. The (former) EU Article 29 Working Party (WP)[5] highlights that **consent** can rarely be a legal ground for processing in an employment setting, due to it not being freely given (A29 WP Report, 2017). Other grounds, such as **legitimate interests** of the employer (such as to monitor for fraudulent employee activity on networks) are possible but need to be balanced against interests of workers whilst also being proportionate and necessary. The most common legal ground for employers processing employee personal data is **necessity** 'for the performance of the employment of contract the data subject is party to… or in order to enter into a contract' (Art 6(1)(b) GDPR). Whilst this is broad and can capture many uses of personal data in the workplace, it is not limitless. The Art 29 WP stresses that data minimisation in any workplace system is important to protect workers, by not storing data longer than needed, and ensuring that any monitoring is transparent and covered by clear workplace policies.

---

[4] https://hudoc.echr.coe.int/eng?i=001-159906
[5] Now the EU Data Protection Board. This older report was still written to reflect GDPR changes, so remains relevant guidance.

One safeguard for workers when employers are conducting high risk data processing activities (which Emotional AI systems like AffecTECH would likely be) is for them to conduct a **data protection impact assessment** (DPIA) (Art 35, GDPR). A DPIA is a valuable tool for demonstrating compliance with GDPR more broadly, as per Art 5(2) of the legislation (see Urquhart, Lodge and Crabtree, 2019). There is some discretion in how impact assessments can be carried out, and some uncertainty persists around when high risk processing occurs (this is the trigger to conduct a DPIA). The UK Information Commissioner Office has usefully helped demystify this for the employment context by providing high risk examples namely: using employee data at scale in profiling e.g. in recruiting and accessing programmes; use of biometrics for access to workplaces or identity verification including fingerprint/voice/facial recognition; tracking location of employees and CCTV monitoring; remote working monitoring; web and cross device tracking; and large scale profiling via wearable based health monitoring.[6] It is clear AC technologies could integrate with a large variety of these systems, and their intrusiveness suggests they would similarly be deemed high risk and require a DPIA.

Another legal development that may help stem some of the issues around AI and AC use in the workplace comes from proposed **EU Directive for Improving Working Conditions in Platform Work** (IWCPL)[7]. This targets specific types of work, namely gig economy platforms such as Uber and Deliveroo, as opposed to workplaces more generally. Given the exploitation these workers often face though, this is a valuable step in protecting them. The proposed law seeks to bring in new mechanisms to help workers deal with **algorithmic tracking**. For example, Article 6 provides for transparency requirements on use of automated monitoring/decision making systems. Here platforms need to tell workers about '**automated monitoring systems'** that are electronically supervising and evaluating their performance. They also need to disclose information about '**automated decision-making systems**' which take decisions 'significantly affecting' the workers conditions. This includes information around the 'categories of actions monitored, supervised and evaluated' not just of the platform itself but also by clients and what 'main parameters' factor into automated decision making processes. It also limits platforms use of personal data of workers where it has to be data 'intrinsically connected to and strictly necessary for the performance of their contract' and cannot include their 'emotional state', their health, psychology, private conversations, and activities during 'off time' (Article 6, IWCPL). This is coupled with provisions for human monitoring of automated systems (Article 7, IWCPL) and **review of significant automated system decisions** when they impact working conditions for workers. This entitles workers to discussion and clarification of the circumstances, reasons and facts around a decision with a 'human contact person at the digital labour platform' (Art 8, IWCPL). This law promises some positive changes and highlights the type of thinking that would benefit other work contexts too (e.g., call centres). We now turn to our data to unpack public concerns and the sorts of issues future legal frameworks may need to address.

---

6 https://ico.org.uk/for-organisations/guide-to-data-protection/guide-to-the-general-data-protection-regulation-gdpr/data-protection-impact-assessments-dpias/examples-of-processing-likely-to-result-in-high-risk/
7 https://ec.europa.eu/social/main.jsp?langId=en&catId=89&newsId=10120&furtherNews=yes

## 5    Data and Analysis: Public Perceptions of AC & Design Fictions

### 5.1    Functionality and accuracy of EAI systems

Most participants were **ambivalent or negative towards our affective computing (AC) use case**, being particularly concerned with the **punitive potential** of the technology and the opportunity for employers to use them as a means to fire staff without engaging with them. Some participants even argued that the systems should only be implemented if they were used to benefit the employee, and that there should be a '*no detriment policy*' (Tom Y2) to their use.

Despite participants' concerns, a range of **benefits** were highlighted, chiefly around the potential to **improve performance**. Some participants felt it could be used as a training tool, helping individuals to '*monitor their performance*' and '*know where to improve*' and the company can then '*help the person to develop the sort of spots where they are lacking*' (Elias Y3). Similarly, it was argued that the system could usefully counter perceived individual failings: '*…if the person that was being evaluated was surly and bad tempered with the customers all the time, and it spots a trend, fine. That person, for want of a better phrase, might need reeducating'* (Cliff D1). Other participants felt it could help with difficult situations and customers, potentially even supporting neurodiverse customers. The older demographic were most positive in this regard, drawing on their own experiences as customers or those having to deal with difficult customers.

Another suggested benefit of the system was that it might contribute to **improvements in workplace culture,** for example by providing a degree of objectivity in evaluation, '*because you're going to cut out any nepotism or any favoritism, because everyone's going to be judged on the same caliber'* (Paul Y2). Impersonal feedback from an affective computer system was also seen as a benefit by one participant in its capacity to reduce interpersonal threat between staff and managers, heightened due to asymmetries in power: '*I think, I would like it better than the supervisors having to check and that and give you that kind of a view. Because this feels like, at least you can kind of question it without feeling like you're undermining their authority or their influence'* (Yasmine E2). Additionally, it might add **competition in a sales environment** or enhance **management opportunities**. These latter elements were dependent on employee characteristics and need to be '*specified to different sectors*', as Alice (Y3) states, it is '*…dependent on what you're ringing up for…in terms of the model employee, you're going to have different personalities for different sectors and different positions.'* She continues to discuss how **context** is key to the value of the system for both employee and customer, noting *'…[if] I'm booking a holiday, that's fine. I don't mind someone being happy and chirpy. If I'm ringing up someone to discuss* *life insurance, you don't want someone on the end of the phone like, "Well, how long do you think you're going to live for?'*

Despite some discussion of benefits, most participants were **negative**, **even hostile**, to the use of emotion sensing tools in the workplace, highlighting their potential to add more **pressure on employees**, be **distracting** and be used **punitively**. Linda (O3) articulates these three points in her criticism of the system: *'...this could possibly cause someone to lose their job. And also, the pressure put on someone, knowing they're being constantly monitored and having little alerts popping up, must be horrendous. I'd hate it...'*. Of even more concern is that it could have **mental health implications** for participants: '*I mean, under the circumstances, you're already stressed with the work. So I think this kind of technology will not help to your mental health.'* (Louis Y3). This anxiety over mental health implications was most pronounced in the groups consisting of disabled participants, both in terms of impact but also how it might contribute to discrimination: '*there's people here that, they've got anxiety, depression, panic attacks, whatever, it doesn't mean they can't sit and do the job but it would put them under even more stress. So that wouldn't be fair even though they could do the job perfectly well.'* (Penny D2). This reinforces the importance of attending to the needs of diverse groups to guard against discrimination of workers with protected characteristics. This is true in terms of research, designing an inclusive approach for collecting data on citizen attitudes, and more broadly when we consider the implementation of new forms of technology.

Participants went on to raise a number of important issues about the implementation of emotion sensing, concerned that it demonstrates a **lack of trust** in employees, with Andy (O1) arguing that employers should, '*[t]rust your employee or get another employee*'. This in turn contributes to **bad morale in the workforce**, who feel '*tense'* because '*You've got the feeling like the bosses can't trust you, so they're putting in a system*' (Samuel Y1). The system itself is criticized as being **highly invasive**, potentially causing you to 'lose people because they would feel that it's too intrusive on their job' (Joanne O2); and potentially open to **manipulation**, which in some cases will be inevitable, demonstrating a lack of trust in existing labour relations: '*It depends who is using the system because then they can manipulate it to do whatever and it will be manipulated in certain situations' (Carol Y3)*. Participants were skeptical of the **effectiveness** of such a system and that it may in fact be detrimental to customers. These negative factors are connected to notions of **accuracy** and participants' view of the **over-simplistic** assumption that the system would be able to interpret the employees', or indeed customers', emotions correctly, failing to account for the prevalence of **individual variation**, that people are different and changeable. As Emily (D2) argues:

*'Everybody's way of going about talking is different. How is it going to be that reliable? Say if you make a joke that's slightly dry humor, sarcastic, the person on the end of the phone, that could be how you should interact with them because that could be them, but then the automatic bot that's recording the calls might not pick up on that. It might pick up on that as being rude. I just think it's not ideal'?*

This key issue underpins a range of concerns around **how the data is interpreted**, the potential for **flaws in the system** and the involvement of different actors leading to **further inaccuracy**. This is illustrated by Cliff's (O1) negative appraisal of the system and AI more generally. He argues that, '*AI is not perfect. It's flawed, as we know*' because '*[s]ystems are built by humans. Humans built in errors and things like that*', which he exemplifies with reference to the 2019 Boeing airplane crashes where, he claims, 'the system didn't react as it was supposed to'.

In its most egregious form, there were anxieties around the potential for **coded bias and discrimination** embedded in these surveillance systems, understood by participants as when the system, '*picks out a certain subgroup that it favours in favour of another subgroup*' (Patrick D2). This discrimination may take the form of unfairly limiting who is employed or negatively impacting an existing employee, leading to employees who '*don't have any personality, they don't have empathy. They don't have anything else apart from just fitting what this machine wants them to.*' (Patrick D2). Ultimately, for many participants, it was seen that the **system needs human intervention** to ensure it worked effectively and as intended and argued that EAI was a poor substitute for an effective manager.

Despite not separating the groups by gender, the data potentially supports Stark et al's (2020) findings on gendered perceptions of workplace surveillance. We found that where there does seem to be divergence in group consensus, it is female participants (Linda O3; Rosie Y2; Brenda D1;) who seem more critical of the technology and/or male participants who present a more positive perspective (Elias Y3; Harry O3; Cliff D1) captured in this interchange from group D1:

> Brenda:    *I wouldn't be happy. I would feel, I think, totally invaded by it. And [partly causing] paranoia, I think.*
> …
> Brenda:    *Feels all negative to me.*
> Cliff:       *I wouldn't say it's all negative.*

A caveat needs to be added that other demographic (and personal) characteristics might be contributing factors. For example, Harry and Elias could be distinguished from other group members based on politics or ethnicity.

### 5.2  Surveillance, worker agency and changing interactions

For some participants, this new emotion sensing system was seen as an **extension of surveillance mechanisms** already in place in workplace settings: *Systems like this, but nowhere near as sophisticated as this, already exist… these days there's software that looks up what you are doing as a worker. I don't know to what extent it measures performance, but it knows exactly what you do at any given time. This is just one stage further on, that's all.* (Tim O1). Signaling the existence of current surveillance technology contributes to a sense of legitimacy or at least resignation towards its implementation (Ball, 2021) in the group discussion. However, if for some participants this 'normalises' the system, for others this makes it even less acceptable as being more **invasive**, continuously present and **lacking human input**. At the most extreme end, it led to participants feeling '*totally invaded*', again contributing to negative mental health implications, such as '*paranoia*' (Brenda D1). This level of surveillance, combined with limitations on an employee's agency and opportunities for interaction, it leads to a workplace environment where employees become '*robots'* (Lauren D2) and dehumanised: '*they don't have any personality, they don't have any empathy. (…) just fitting what this machine wants them to.*' (Patrick D2).

Exacerbating these issues is the **power asymmetries and lack of control** employees have over the implementation, use and functionality of these systems. This leads to fears around how they would be used in automating management structures, and employee evaluation. Phillip (O1) articulates this well stating it may '*replace reviews and appraisals and that sort of thing. You could imagine, the option is there to be sacked automatically by email, without even speaking to anybody. And that would be a bit of a concern, I think. It's about, yeah, where it ends up.*' Similarly (Penny D2) states employees only option could be quitting if they do not like these systems.

## 5. Conclusions

We have explored the wider context of legal concerns and historical issues of incorporating new IT, like AC, into the workplace. By using our innovative design fiction approach to collecting citizen attitudes, this research has revealed a number of important concerns around implementation of AC in the workplace. While most participants were **ambivalent or negative towards** the use case, a number of **benefits** were suggested. These focused on the technologies use for **training to improve performance** and potential to **enhance workplace culture**. However, these were both seen to be contingent on **context**, in terms of the individual and sector.

The pre-existence of a range of **surveillance mechanisms in the workplace legitimised** the implementation of AC systems for some participants, while others found themselves **resigned** to their use. Most participants were **strongly negative** towards the use of AC systems in the workplace, pointing to the **pressure** they put on employees and their potential to be **distracting** and used **punitively**, with **mental health implications**. This latter was most strongly seen in groups self-identifying as disabled, highlighting the importance of this research that collected citizen attitudes from groups who have traditionally been ignored in the development of new technology, the implementation of which may reinforce existing inequalities.

Participants argued that these surveillance systems contribute to **bad morale in the workforce**, highlighting a **lack of trust** in employees. The systems are seen as both **highly invasive** and **open to manipulation** while at the same time **in-effective** and **inaccurate** relying on **over-simplistic** interpretations that is unable to account for **individual variation**. Participants raised legitimate concerns about **systemic issues**, including **coded bias and discrimination** as well as the dangerous potential for **human error**. As with many of the other use-cases discussed in the workshop, participants were keen for some element of **human intervention** to support or balance out decisions made by AI systems.

A key concern emerging from this research is the reinforcement of awareness that these workplace uses of AC are being introduced in contexts with existing **power asymmetries** and they have great potential to magnify inequalities between employers and employees; this is something citizens understand and are anxious about, seeing the potential for the negative impact on labour relations and individual wellbeing. These concerns must be attended to in the development, implementation and legislation surrounding AC deployment in the workplace.